\documentclass[a4paper,11pt]{article}

\usepackage{jheppub} 

\usepackage[T1]{fontenc} 
\usepackage[dvipsnames]{xcolor}

\usepackage{makeidx}
\makeindex
\usepackage{amstext}
\usepackage{hyperref}
\usepackage{amsmath}
\usepackage{amsfonts}
\usepackage{bbding}
\usepackage{amssymb}
\usepackage{color}
\usepackage{cancel}
\usepackage{slashed}
\usepackage{empheq} 

\usepackage{graphicx}
\usepackage{mathtools} 
\usepackage{stackrel}
\usepackage{multirow}
\usepackage{makecell}
\usepackage{mathrsfs}

\usepackage{arydshln} 

\usepackage{fancyvrb}

\usepackage{hhline}

\usepackage{enumitem}

\vbadness=\maxdimen

\usepackage{slashed} 

\allowdisplaybreaks

\usepackage[normalem]{ulem}
\usepackage{wasysym}

\usepackage{simplewick}
\usepackage{pifont}
\usepackage{varioref}

\usepackage{float} 

\usepackage{datetime}

\title{\boldmath 
The Exact Amplitudes of Six Polarization Modes for Gravitational Waves}

\author[a]{Young-Hwan Hyun}
\author[b]{, Yoonbai Kim}
\author[b]{, Seokcheon Lee}

\affiliation[a]{Korea Institute of Science and Technology Information (KISTI), 245 Daehak-ro, Yuseong-gu, Daejeon 34141, Republic of Korea}
\affiliation[b]{Department of Physics and Institute of Basic Science,
Sungkyunkwan University, Suwon 16419, Republic of Korea}

\emailAdd{younghwan.hyun@gmail.com}
\emailAdd{yoonbai@skku.edu}
\emailAdd{skylee2@gmail.com}

\abstract{The formalism of the exact six polarization modes of gravitational waves is constructed in terms of both the small metric perturbations and the Newman-Penrose scalars. The obtained formulae are applicable to any metric-compatible gravity theories whose gravitational waves propagate along either the null or non-null geodesics. Once a gravity theory, specifically its linearized wave equation, is written, comparison to the observed data of the laser interferometer experiments is direct.} 
\begin{document}

\maketitle
\flushbottom

\section{Introduction}
\label{sec:intro}

To date, the Einstein's general relativity (GR) has passed all tests since it was developed in 1916, and thus it is important to come up with an extension of gravity theory that is allowed in these same tests. Such longevity is not only related to its absolute correctness, but can also motivate more accurate tests to probe the correction to the Einstein's GR. New precession searches for small deviations from GR is intriguing in the context of astrophysics and cosmology. The first candidate experiment for identifying the violation of GR is to look for the possible polarization modes of gravitational waves (GWs) and its formulation was constructed firstly in Ref. \cite{Eardley:1974nw,Eardley:1973br} (see also reviews~\cite{Will:1993ns,Will:2014kxa}).\\

Since Einstein's GR predicted the existence of gravitational waves~\cite{Einstein:1916cc}, a long-awaited signal of gravitational waves was picked up by the advanced Laser Interferometer Gravitational-Wave Observatory (aLIGO) and Virgo collaboration \cite{Abbott:2016blz,Abbott:2016nmj,TheLIGOScientific:2016pea,Abbott:2017vtc}. This milestone in gravitational wave research opens a window to probe the highly dynamical and strong-field regimes of gravity \cite{Rizwana:2016qdq,Berti:2015itd}. In addition, the aLIGO and Virgo also allow the precision study of the polarization modes of the gravitational waves, particularly the bound of the non-tensorial modes \cite{Abbott:2017tlp,Abbott:2018utx}. The analysis for known galactic pulsars put the constraint on the strain of the scalar and vector modes to be below $1.5\times 10^{-26}$ at 95\% credibility~\cite{Abbott:2018utx}, which is the first direct upper limit for a non-tensorial strain. This upper bound provides a guideline to modify the beyond-GR theories of gravity.\\

In the context of metric-compatible theories, the polarization modes consist of six modes which are called breathing ($b$), longitudinal ($l$), vector-$x$ ($x$), vector-$y$ ($y$), plus ($+$), and cross $(\times $) modes. The Einstein's GR predicts transverse and traceless waves whose quantization leads to massless spin-two gravitons and thus the sole detection of the two tensor modes, plus and cross polarization modes, will fulfill the GR's prediction. The systematic study of the six polarization modes of gravitational waves has been made under the assumption of the weak, plane, and null propagation and analyzed in terms of the Newman-Penrose (NP) formalism~\cite{Eardley:1974nw}. Most of the subsequent researches on various extended models of gravity have employed this formalism with E(2) classification to calculate the NP scalars corresponding to each polarization mode
~\cite{Nishizawa:2009bf,Alves:2009eg,Rizwana:2016qdq,Myung:2016zdl},
even on the theories involving massive modes
~\cite{Bessada:2009qw,dePaula:2004bc,Rizwana:2016qdq,Abedi:2017jqx,Alves:2010ms}.
In the case of the bi-metric theory, it has been shown by the use of the NP scalars that how massive degrees of freedom contribute to the amplitude of non-tentorial modes \cite{dePaula:2004bc,Corda:2007zz}. This is because the NP scalars provide the simplest way to look into a specific propagation of gravitational waves even in extended gravity theories, however the NP analysis in Ref.~\cite{Eardley:1974nw} is not exact for the massive gravity theories anymore. Therefore it is necessary to construct the exact formalism for the six polarization modes of the non-null propagating gravitational waves. Recently, this point was indicated in 
Ref.~\cite{Liang:2017ahj}.\\

It is timely to reconstruct the formalism to give a correct interpretation for the non-null propagations from the observed data of the gravitational waves. In this work, we obtain the formulae of the six polarization amplitudes connecting the observed data from the laser interferometers and the GWs of the proposed gravity theory. These are also applicable to the non-null propagation of the GWs.
Let us begin with introducing the assumptions of our formalism:
\begin{itemize}[noitemsep]
\item[1.] The gravity theories of consideration are metric-compatible.
\item[2.] There exists the weak gravity limit in which the gravitational waves are governed by linear wave equations. This is the so-called short wavelength approximation.
\end{itemize}
The aforementioned assumptions dictate the following guidelines:
\begin{itemize}[noitemsep]
\item[1.] Since any metric-compatible theory is allowed, the geodesic equation and the Bianchi identity can be used. On the other hand, the specific form of the action, e.g. the Einstein-Hilbert action for GR, or equivalently the corresponding dynamical equations, e.g. the Einstein equations, need not be assumed in a derivation of the formalism. In application, it means that  any metric-compatible gravity action, which can involve not only GR but also many other candidate theories, e.g., higher derivative or $f(R)$ or massive  gravity theories, can utilize our formalism without restriction.
\item[2.] The linear wave equations for the weak gravitation field $h_{\mu\nu}$ let physical contents of the GWs be read through the dispersion relation $\omega=\omega(\boldsymbol{k})$ and their six polarization modes. 
\end{itemize}
The six polarization modes are formulated in terms of both the NP scalars and the six physical degrees among ten components of $h_{\mu\nu}$ by appropriate gauge fixing. Since the formalism is written up to the response function, a comparison between the theory, say the action, and the observed data can directly be performed.\\ 

This work is organized as follows. In section~\ref{Sec2}, the formalism in Ref.~\cite{Eardley:1974nw} is reviewed. In subsection~\ref{Sec3.1}, we explain the description of six polarization modes based on the usual NP formalism. We express the exact driving-force matrix for the plane-wave weak propagations of gravitational waves based on the NP formalism in subsection~\ref{Sec3.2} and in terms of the metric perturbations in subsection~\ref{Sec3.3}. Discussion on the difference between the usual and exact results is also accompanied. In subsection~\ref{Sec3.4}, the response functions are obtained. Some known gravity models are analyzed in section~\ref{Sec4}. We conclude in section~\ref{Conclusion} with a few research directions.

\section{Six observables of gravitational waves}\label{Sec2}

When a freely falling observer is at a fiducial point in an approximately Lorentz normal coordinate system $(t,x^{i})=(t,x,y,z)$ for the spatial coordinate $x^{i}$ of the test particle at rest, the acceleration relative to the location of the observer is depicted by the geodesic deviation equation \cite{Eardley:1974nw},
\begin{align}
a_{i}=-R_{0i0j}x^{j}\,,
\end{align}
where the electric components of the Riemann tensor $R_{0i0j}$, the so-called Riemann field, is the only measurable quantity in the gravitational wave detection. Suppose that propagating gravitational wave is weak and a plane-wave. When the $z$-direction is chosen parallel to the propagation of gravitational waves,  every component of the Riemann field $R_{0i0j}(t_{{\rm r}})$ becomes a function of a retarded time, $t_{{\rm r}}=t-z/v$. \\

The total six electric components of the Riemann tensor are set by the symmetric driving-force matrix $S_{ij}(t)$ \cite{Eardley:1974nw,Will:2005va},
\begin{align}
\label{DFdef}
S_{ij}(t_{{\rm r}})\equiv R_{0i0j}(t_{{\rm r}})\,.
\end{align}
Since this driving-force matrix possesses six independent degrees, the six basis polarization matrices are introduced as
\begin{align}
&E_{1}(\hat{z})=
\begin{pmatrix}
0 & 0 & 0 \\
0 & 0 & 0 \\
0 & 0 & 1 \\
\end{pmatrix},~~
~~E_{2}(\hat{z})=
\begin{pmatrix}
0 & 0 & 1 \\
0 & 0 & 0 \\
1 & 0 & 0 \\
\end{pmatrix},~~
E_{3}(\hat{z})=
\begin{pmatrix}
0 & 0 & 0 \\
0 & 0 & 1 \\
0 & 1 & 0 \\
\end{pmatrix},~~\nonumber\\
&E_{4}(\hat{z})=\frac{1}{2}
\begin{pmatrix}
1 & 0 & 0 \\
0 & -1 & 0 \\
0 & 0 & 0 \\
\end{pmatrix},~~
E_{5}(\hat{z})=
\begin{pmatrix}
0 & 1 & 0 \\
1 & 0 & 0 \\
0 & 0 & 0 \\
\end{pmatrix},~~
E_{6}(\hat{z})=\frac{1}{2}
\begin{pmatrix}
1 & 0 & 0 \\
0 & 1 & 0 \\
0 & 0 & 0 \\
\end{pmatrix}\,.
\label{basis_matrix}
\end{align}
Note that the coefficients in front of the matrices were set differently in Ref.~\cite{Eardley:1974nw} to read the polarization amplitudes in the Newman-Penrose (NP) formalism.
In the basis of polarization matrices, the driving-force matrix is expanded with polarization amplitudes $p_{n}$,
\begin{align}
S(t)=\sum_{A=1}^{6}p_{A}(\hat{z},t)E_{A}(\hat{z})\,,
\label{Sdef}
\end{align}
and comparison of \eqref{DFdef} and \eqref{Sdef} gives 
\begin{align}
S=
\begin{pmatrix}
R_{txtx} & R_{txty} & R_{txtz} \\
R_{tytx} & R_{tyty} & R_{tytz} \\
R_{tztx} & R_{tzty} & R_{tztz} \\
\end{pmatrix}
=
\begin{pmatrix}
\frac{1}{2}(p_{4}+p_{6}) & p_{5} & p_{2} \\
p_{5} & \frac{1}{2}(-p_{4}+p_{6}) & p_{3} \\
p_{2} & p_{3} & p_{1} \\
\end{pmatrix}\,.
\label{DFmatrix}
\end{align}
Each polarization amplitude of the six electric components, $p_{1}, \cdots, p_{6}$, corresponds to a specific geometrical distortion of the test particle distribution, whose  
\begin{figure}[h]
\centering
\includegraphics[scale=0.4]{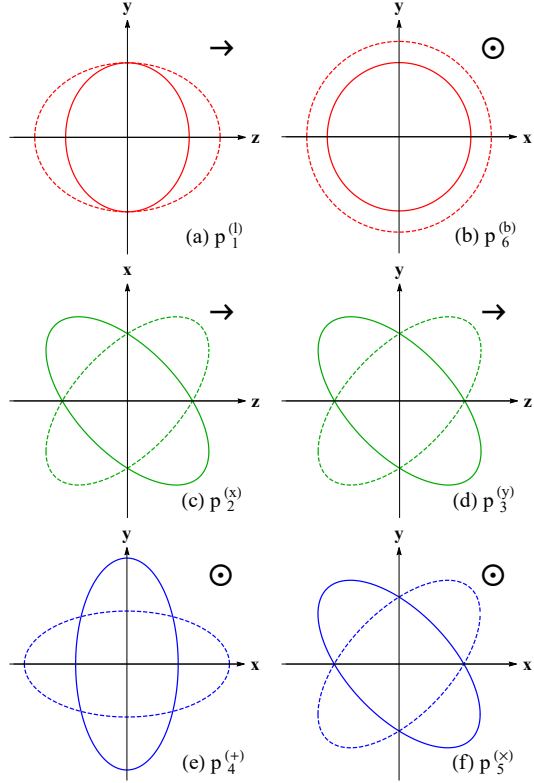}
\caption{Six polarization modes: (a) breathing mode $p_{1}^{(b)}$, (b) longitudinal mode $p_{6}^{(l)}$, (c) vector-$x$ mode $p_{2}^{(x)}$, (d) vector-$y$ mode $p_{3}^{(y)}$, (e) plus mode $p_{4}^{(+)}$, (f) cross mode $p_{5}^{(\times)}$. Here we added the superscript of every corresponding polarization mode to $p_{n}$ to show clearly its geometrical description. The red, green, and blue colors indicate scalar, vector, and tensor modes, respectively. The circled dot in (b), (e), and (f) indicates respectively the wave propagating out of the page,
and the right arrows in (a), (c), and (d) mean respectively the direction of wave propagation in the $z$-direction.}
\label{6modes}
\end{figure}
shapes are displayed in Fig.~\ref{6modes} (See \cite{Eardley:1974nw}). The name of each mode $p_{1},\cdots ,p_{6}$ is the longitudinal, vector-$x$, vector-$y$, plus, cross, breathing polarization mode, respectively. Thus, in our basis \eqref{basis_matrix}, the exact polarization amplitudes are written in terms of the driving-force matrix element in a simplest form,\begin{align}
&~p_{1}^{(l)}\equiv  R_{tztz},~~~~~~~~~~~~~~~p_{2}^{(x)}\equiv  R_{tztx},~~~~~~~~~p_{3}^{(y)}\equiv  R_{tzty},\nonumber\\
&p_{4}^{(+)}\equiv  R_{txtx}-R_{tyty},~~~~p_{5}^{(\times )}\equiv  R_{txty},~~~~~~~~~p_{6}^{(b)}\equiv  R_{txtx}+R_{tyty}\,,
\label{pn}
\end{align}
where we add the description of the mode in the superscript for a clear distinction. 

\section{Polarization modes}\label{Sec3}

\subsection{Null propagation of gravitational waves}
\label{Sec3.1}

In this subsection, we briefly recapitulate the traditional method on the six polarization modes of the massless gravitons in which the null propagation assumption is adopted~\cite{Eardley:1974nw}. We will examine the amplitude expressions in  the traditional NP method to find
necessary corrections to extend the exact formalism to the massive gravitational waves tracking the non-null geodesic.\\

For the description of the polarization modes under the null propagation assumption, it is convenient to introduce the NP quantities for simplicity. For a local null tetrad basis, $k$, and two null spin tetrad, $m$, and $\bar{m}$, we have the four tetrad basis vectors,
\begin{align}
\boldsymbol{k}&=\frac{1}{\sqrt{2}}(\boldsymbol{\partial}_{t}+\boldsymbol{\partial}_{z}),~~~~~~~\,\boldsymbol{l}=\frac{1}{\sqrt{2}}(\boldsymbol{\partial}_{t}-\boldsymbol{\partial}_{z}),\nonumber\\
\boldsymbol{m}&=\frac{1}{\sqrt{2}}(\boldsymbol{\partial}_{x}+i\boldsymbol{\partial}_{y}),~~~~\boldsymbol{\bar{m}}=\frac{1}{\sqrt{2}}(\boldsymbol{\partial}_{x}-i\boldsymbol{\partial}_{y}),
\end{align}
which satisfy the normalization conditions\begin{align}
k_{\mu}l^{\mu}=-1,~~~~m_{\mu}\bar{m}^{\mu}=1\,.
\end{align}
In four dimensions, the Riemann tensor is split into three irreducible parts, $C_{\mu\nu\rho\sigma}$, $R_{\mu\nu}-\frac{1}{4}g_{\mu\nu}R$, and $R$, where the Weyl tensor in the four-dimensional spacetime is defined by 
\begin{align}
C_{\mu\nu\rho\sigma}
=R_{\mu\nu\rho\sigma}-2g_{[\mu|[\rho}R_{\sigma]|\nu]}
+\frac{1}{3}g_{\mu[\rho}g_{\sigma]\nu}R
.
\end{align} 
In the NP formalism, the five complex Weyl-NP scalars are defined and classified with spin-weights from the Weyl tensor
\begin{align}\label{NPWeyl}
s=+2:~\Psi_{0}&\equiv C_{kmkm}~,
        \nonumber\\
s=+1:~\Psi_{1}&\equiv C_{klkm}=C_{\bar{m}mkm}~,
        \nonumber\\
s=0~~\,:~\Psi_{2}&\equiv C_{km\bar{m}l}=\frac{1}{2}(C_{klkl}+C_{kl\bar{m}m})
        =\frac{1}{2}(C_{\bar{m}m\bar{m}m}+C_{kl\bar{m}m})~,
        \nonumber\\
s=-1:~\Psi_{3}&\equiv C_{kl\bar{m}l}=C_{\bar{m}m\bar{m}l}~,
        \nonumber\\
s=-2:~\Psi_{4}&\equiv C_{\bar{m}l\bar{m}l}~.
\end{align}
The ten Ricci-NP scalars are defined from the traceless and trace parts of the Ricci tensor $R_{\mu\nu}$ as
\begingroup
\allowdisplaybreaks
\begin{align}\label{NPRicci}
s=+2&:~~~~\Phi_{02}\equiv \frac{1}{2}R_{mm} \, , \nonumber\\
s=+1&:~
\begin{cases}
\displaystyle\Phi_{01}\equiv \frac{1}{2}R_{km} \, ,\\
\displaystyle\Phi_{12}\equiv \frac{1}{2}R_{lm} \, , \\
\end{cases} \nonumber\\
s=0~~\,&:~
\begin{cases}
\displaystyle\Phi_{00}\equiv \frac{1}{2}R_{kk}  \, , \\
\displaystyle \Phi_{11}\equiv\frac{1}{4}(R_{kl}+R_{m\bar{m}}) \, , \\
\displaystyle\Phi_{22}\equiv \frac{1}{2}R_{ll} \, ,  \\
\end{cases}\nonumber\\
s=-1&:~
\begin{cases}
\displaystyle\Phi_{10}\equiv \frac{1}{2}R_{k\bar{m}} =\Phi_{01}^{*}\, ,  \\
\displaystyle\Phi_{21}\equiv \frac{1}{2} R_{l\bar{m}} =\Phi_{12}^{*}\, , \\
\end{cases} \nonumber\\
s=-2&:~~~~\Phi_{20}\equiv \frac{1}{2}R_{\bar{m}\bar{m}} =\Phi_{02}^{*}\, ,  \nonumber\\
&~~~~~~~~~\Lambda\equiv \frac{R}{24}=\frac{1}{12}(R_{m\bar{m}}-R_{kl}) \, .
\end{align}
\endgroup 
Under the null condition the measurable field becomes a function of the retarded time $t_{{\rm r}}=t-z$ with $v=1$, and thus the Riemann tensor satisfies
\begin{align}
R_{abcd,p}=0\,,
\label{PC}
\end{align}
where $(a,b,c,d)$ range over $(k,l,m,\bar{m})$ and $(p,q,ρ\cdots )$ range only over $(k,m,\bar{m})$. With the help of the Bianchi identity, 
\begin{align}
R_{ab[pq,l]}=\frac{1}{3}(R_{abpq,l}+R_{abql,p}+R_{ablp,q})=\frac{1}{3}R_{abpq,l}=0\,,
\end{align}
the equation~\eqref{PC} leads to a constant curvature solution. Since any non-vanishing constant curvature solution is irrelevant to wave phenomenon, only the solution of our interest should have a vanishing Riemann tensor component,
\begin{align}
R_{abpq}=0=R_{pqab}\, .
\label{N}
\end{align}
Therefore, all non-vanishing components of the Riemann tensor should take the form $R_{plql}$. Accordingly, under the null condition, all the NP scalars in \eqref{NPWeyl} and \eqref{NPRicci} are given by
\begingroup
\allowdisplaybreaks
\begin{align}
&\Psi_{0}= C_{kmkm}=R_{kmkm}\xrightarrow{\text{\small null}}0\,,\nonumber\\
&\Psi_{1}=C_{klkm}=R_{klkm}-\frac{1}{2}R_{km}\xrightarrow{\text{\small null}}0~,\nonumber\\
&\Psi_{2}= C_{km\bar{m}l}=R_{km\bar{m}l}-\frac{1}{12}R\xrightarrow{\text{\small null}}\frac{1}{6}R_{klkl}~,\nonumber\\
&\Psi_{3}=C_{kl\bar{m}l}=R_{kl\bar{m}l}-\frac{1}{2}R_{l\bar{m}}\xrightarrow{\text{\small null}}\frac{1}{2}R_{kl\bar{m}l}~,\nonumber\\
&\Psi_{4}=C_{\bar{m}l\bar{m}l}=R_{\bar{m}l\bar{m}l}\xrightarrow{\text{\small null}} R_{\bar{m}l\bar{m}l},\nonumber\\
&\Phi_{00}=\frac{1}{2}R_{kk}\xrightarrow{\text{\small null}}0,\nonumber\\
&\Phi_{01}=\Phi_{10}^{*}=\frac{1}{2}R_{km}\xrightarrow{\text{\small null}}0,\nonumber\\
&\Phi_{02}=\Phi_{20}^{*}=\frac{1}{2}R_{mm}\xrightarrow{\text{\small null}}0,\nonumber\\
&\Phi_{11}=\frac{1}{4}(R_{kl}+R_{m\bar{m}})\xrightarrow{\text{\small null}}\frac{1}{4}R_{klkl}=\frac{3}{2}\Psi_{2}(=\Psi_{2}-\Lambda)~,\nonumber\\
&\Phi_{12}=\Phi_{21}^{*}=\frac{1}{2}R_{lm}\xrightarrow{\text{\small null}}\frac{1}{2}R_{klml}=\Psi_{3}^{*}\nonumber\\
&\Phi_{22}=\frac{1}{2}R_{ll}=R_{ml\bar{m}l}\xrightarrow{\text{\small null}}R_{ml\bar{m}l}~,\nonumber\\
&\Lambda=\frac{R}{24}=-\frac{1}{12}(R_{kl}-R_{m\bar{m}})\xrightarrow{\text{\small null}}-\frac{1}{12}R_{klkl}=-\frac{1}{2}\Psi_{2}\, ,
\label{NPnull_relation}
\end{align}
\endgroup 
where $R=-2R_{kl}=-2R_{klkl}$ is used in the last formula. The eight NP scalars among all the fifteen NP scalars do not vanish, but only four NP scalars, $\Psi_{2},~\Psi_{3},~\Psi_{4},~\Phi_{22}$, correspond to independent components of the Riemann tensor. We shall call these four NP scalars ``NP-null scalars''. Since $\Psi_{2},\Phi_{22}$ are real and $\Psi_{3},\Psi_{4}$ are complex in~\eqref{NPnull_relation} by applying the null condition, the NP-null scalars have six real degrees as shown in the Table below:   
\begin{table}[H]
\begin{center}
\begin{tabular}{|c|c|c|}\hhline{-~-}
NP scalars &  & NP-null scalars \\\hhline{=~=}
$\Psi_{0},~\Psi_{1},~\Psi_{2},~\Psi_{3},~\Psi_{4}$ &  & \\\hhline{~~~}
$\Phi_{02},~\Phi_{12},~\Phi_{01},
~\Phi_{00},~\Phi_{11},~\Phi_{22},~\Phi_{10},~\Phi_{21},~\Phi_{20}$ 
& $\xrightarrow{\text{\tiny null condition}}$  
& $\Psi_{2},~\Psi_{3},~\Psi_{4},~\Phi_{22}$
\\\hhline{~~~}
$\Lambda$ &  & \\\hhline{-~-}
\end{tabular}
\end{center}
\end{table}
These six real degrees of the NP-null scalars correspond to the polarization amplitude, $p_{n}$ by~\eqref{pn},
\begingroup
\allowdisplaybreaks
\begin{align}
\Psi_{2}&\xrightarrow{\text{\small null}}\frac{1}{6}R_{lklk}=\frac{1}{6}R_{tztz}~~~~~~~~~~~~~~~~~~~\equiv \frac{1}{6}p_{1}^{(l)}({\vec k},t) \, , \nonumber\\
{\rm Re}(\Psi_{3})&\xrightarrow{\text{\small null}}\frac{1}{2}{\rm Re}(R_{lkl\bar{m}})\xrightarrow{\text{\small null}}\frac{1}{2}R_{tztx}~~~~~~~~\equiv \frac{1}{2}p_{2}^{(x)}({\vec k},t)\,,\nonumber\\
{\rm Im}(\Psi_{3})&\xrightarrow{\text{\small null}}\frac{1}{2}{\rm Im}(R_{lkl\bar{m}})\xrightarrow{\text{\small null}}-\frac{1}{2}R_{tzty}~~~~~~\equiv-\frac{1}{2} p_{3}^{(y)}({\vec k},t)\,,\nonumber\\
{\rm Re}(\Psi_{4})&\xrightarrow{\text{\small null}}{\rm Re}(R_{l\bar{m}l\bar{m}})~\xrightarrow{\text{\small null}}R_{txtx}-R_{tyty}\,\equiv p_{4}^{(+)}({\vec k},t)\,,\nonumber\\
{\rm Im(}\Psi_{4})&\xrightarrow{\text{\small null}}{\rm Im}(R_{l\bar{m}l\bar{m}})~\xrightarrow{\text{\small null}}-2R_{txty}~~~~~~~\equiv -2p_{5}^{(\times )}({\vec k},t)\,,\nonumber\\
\Phi_{22}&\xrightarrow{\text{\small null}}R_{lml\bar{m}}~~~~~~~\xrightarrow{\text{\small null}}R_{txtx}+R_{tyty}\,\equiv p_{6}^{(b)}({\vec k},t)\,,
\label{pnbar}
\end{align}
and the driving-force matrix~\eqref{DFmatrix} is written under the null-propagation condition in terms of the NP-null scalars as
\begin{align}
S_{{\textrm{null}}}=\begin{pmatrix}
\frac{1}{2}[{\rm Re}(\Psi_{4})+\Phi_{22})] & -\frac{1}{2}
{\rm Im(}\Psi_{4}) & 2
{\rm Re}(\Psi_{3}) \\
-\frac{1}{2}{\rm Im(}\Psi_{4}) & -\frac{1}{2}[{\rm Re}(\Psi_{4})-\Phi_{22}] & -2{\rm Im}(\Psi_{3}) \\
2{\rm Re}(\Psi_{3}) & -2{\rm Im}(\Psi_{3}) & 6\Psi_{2} \\
\end{pmatrix}\,.
\label{S}
\end{align}
\endgroup 

The polarization amplitude $p_{n}$ in~\eqref{pnbar} is different from that in Ref.~\cite{Eardley:1974nw}. First, the overall sign in \eqref{pnbar} is opposite to that in Ref.~\cite{Eardley:1974nw} since we used the definition of the NP scalars in Ref.~\cite{Frolov:1998wf}. Second, each $p_{n}$ in \eqref{pnbar} has a different coefficient since the basis polarization matrices in \eqref{basis_matrix} chose different normalization coefficients. If six normalization coefficients $a_{n}$'s are introduced as
\begingroup
\allowdisplaybreaks
\begin{align}
&E_{1}(\hat{z})=a_{1}
\begin{pmatrix}
0 & 0 & 0 \\
0 & 0 & 0 \\
0 & 0 & 1 \\
\end{pmatrix},~~
~~E_{2}(\hat{z})=a_{2}
\begin{pmatrix}
0 & 0 & 1 \\
0 & 0 & 0 \\
1 & 0 & 0 \\
\end{pmatrix},~~
E_{3}(\hat{z})=a_{3}
\begin{pmatrix}
0 & 0 & 0 \\
0 & 0 & 1 \\
0 & 1 & 0 \\
\end{pmatrix},~~\nonumber\\
&E_{4}(\hat{z})=a_{4}
\begin{pmatrix}
1 & 0 & 0 \\
0 & -1 & 0 \\
0 & 0 & 0 \\
\end{pmatrix},~~
E_{5}(\hat{z})=a_{5}
\begin{pmatrix}
0 & 1 & 0 \\
1 & 0 & 0 \\
0 & 0 & 0 \\
\end{pmatrix},~~
E_{6}(\hat{z})=a_{6}
\begin{pmatrix}
1 & 0 & 0 \\
0 & 1 & 0 \\
0 & 0 & 0 \\
\end{pmatrix}\,,
\label{abasis_matrix}
\end{align} 
\endgroup 
then they have different values as in the Table below:
\begin{center}  
\begin{tabular}{|c|cccccc|}\hline
 & $a_{1}$ & $a_{2}$ & $a_{3}$ & $a_{4}$ & $a_{5}$ & $a_{6}$ \\\hline
\eqref{basis_matrix} & 1 & 1 & 1 & $\frac{1}{2}$ & 1 & $\frac{1}{2}$ \\\hline
Ref.~\cite{Eardley:1974nw} & $-6$ & $-2$ & 2 & $-\frac{1}{2}$ & $\frac{1}{2}$ & $-\frac{1}{2}$ \\\hline
\end{tabular}
\end{center}
Subsequently, the polarization amplitude $p_{n}$ in \eqref{pnbar} is related to the corresponding amplitude $\bar{p}_{n}$ in \cite{Eardley:1974nw},
\begin{align}
&p_{1}^{(l)}= -6\bar{p}_{1}^{(l)},~~~~~~~p_{2}^{(x)}= -2\bar{p}_{2}^{(x)},
~~~~~~~p_{3}^{(y)}= 2\bar{p}_{3}^{(y)},~~\nonumber\\
&p_{4}^{(+)}=-\bar{p}_{4}^{(+)},
~~~~~~p_{5}^{(\times )}=\frac{1}{2}\bar{p}_{5}^{(\times )},
\,~~~~~~~~p_{6}^{(b)}=-\bar{p}_{6}^{(b)}
\, .
\end{align}
The driving-force matrix $S_{\text{null}}$ for the null condition in \eqref{S}, a physical quantity, coincides exactly irrespective of the choice of the normalization constants in \eqref{abasis_matrix}. 
\vspace{12pt}

\subsection{Non-null propagation of gravitational waves in terms of NP scalars}\label{Sec3.2}

The gravitational waves generated by some gravitational theories may propagate along non-null geodesics. Since the NP formalism \eqref{pnbar} obtained under the null condition \eqref{N} cannot be applied anymore to those, it is necessary to find the six polarization amplitudes $p_{n}~(p=1,2,\cdots,6)$ before assigning the null condition. The exact polarization amplitudes expressed in terms of the electric components of the Riemann tensor are easily obtained by inverting the five complex Weyl-NP scalars \eqref{NPWeyl} and the ten Ricci-NP scalars \eqref{NPRicci},
\begin{align}
p_{1}^{(l)}&=R_{tztz}~~~~~~~~~~\, 
=2[{\textrm{Re}}(\Psi_{2})+\Phi_{11}-\Lambda]  
\, ,
\nonumber\\
p_{2}^{(x)} 
& =R_{tztx}~~~~~~~~~~\,  
=-{\textrm{Re}}(\Psi_{1})+{\textrm{Re}}(\Psi_{3})-{\textrm{Re}}(\Phi_{01})
+{\textrm{Re}}(\Phi_{12}) 
\, , 
\nonumber\\
p_{3}^{(y)} 
& =R_{tzty} ~~~~~~~~~~\, 
=-{\textrm{Im}}(\Psi_{1})-{\textrm{Im}}(\Psi_{3})-{\textrm{Im}}(\Phi_{01})
+{\textrm{Im}}(\Phi_{12})  
\, ,
\nonumber\\
p_{4}^{(+)} 
& =R_{txtx}-R_{tyty}  
={\textrm{Re}}(\Psi_{0})+{\textrm{Re}}(\Psi_{4})-2{\textrm{Re}}(\Phi_{02})
\, ,
\nonumber\\
p_{5}^{(\times )} 
& =R_{txty}~~~~~~~~~~\,  
={\frac{1}{2}[\textrm{Im}}(\Psi_{0})-{\textrm{Im}}(\Psi_{4})
-2{\textrm{Im}}(\Phi_{02})] 
\, ,
\nonumber\\
p_{6}^{(b)} 
& =R_{txtx}+R_{tyty}  
=-2{\textrm{Re}}(\Psi_{2})+\Phi_{00}+\Phi_{22}-4\Lambda
\, .
\label{6}
\end{align}

The exact NP expressions valid for plane-wave amplitudes of gravitational waves are obtained by assigning the condition of the plane-wave propagation along the $z$-direction to the components of the Riemann tensor. Specifically, every component of the Riemann tensor for the plane-wave is a function of time $t$ and propagation coordinate $z$ including  the retarded time  with $v$, $t_{{\textrm{r}}}=t-z/v$, 
$R_{\mu\nu\rho\sigma}=R_{\mu\nu\rho\sigma}(t,z)$, which satisfies
\begin{align}
R_{\mu\nu\rho\sigma,p}=0\,,
\end{align} 
where $v$ is the speed of the gravitational wave of consideration, and $(\mu,\nu,\rho,\sigma)$ range over $(t,x,y,z)$ and $(p,q,r,\cdots )$ range only over $(x,y)$. Except for a trivial non-wavelike constant solutions of no interest, the Bianchi identity, $R_{\mu\nu[pq,t]}=0=\frac{1}{3}R_{\mu\nu pq,t}$, supports some null curvature solutions for the gravitational waves, 
\begin{align}
R_{\mu\nu pq}=0\,.
\label{planecondition}
\end{align}
Since the Ricci and Einstein tensors are related to the polarization amplitudes as
\begin{align}
p_{1}^{(l)}&=\frac{1}{2}(G_{tt}+G_{xx}+G_{yy}-G_{zz})-R_{xyxy}\,\xrightarrow[\text{wave}]{\text{plane}}\frac{1}{2}(G_{tt}+G_{xx}+G_{yy}-G_{zz})\,, 
\nonumber\\
p_{2}^{(x)}&=-G_{xz}+R_{zyxy}~~~~~~~~~~~~~~~~~~~~~~~~~~~~\xrightarrow[\text{wave}]{\text{plane}} -G_{xz}\,,  
\nonumber\\
p_{3}^{(y)}&=-G_{yz}-R_{zxxy}~~~~~~~~~~~~~~~~~~~~~~~~~~~~\xrightarrow[\text{wave}]{\text{plane}}-G_{yz}\,, 
\nonumber\\
p_{4}^{(+)}&=-(G_{xx}-G_{yy})+R_{zxzx}-R_{zyzy}~~~~~~\xrightarrow[\text{wave}]{\text{plane}}-(G_{xx}-G_{yy})+R_{zxzx}-R_{zyzy},\nonumber\\
p_{5}^{(\times )}&=-G_{xy}+R_{zxzy}~~~~~~~~~~~~~~~~~~~~~~~~~~~~\xrightarrow[\text{wave}]{\text{plane}}-G_{xy}+R_{zxzy}\,,\nonumber\\
p_{6}^{(b)}&=G_{zz}+R_{xyxy}~~~~~~~~~~~~~~~~~~~~~~~~~~~~~~\,\xrightarrow[\text{wave}]{\text{plane}}G_{zz}\,,  
\label{pwithRicci}
\end{align}
This plane-wave condition \eqref{planecondition} enables us to read easily vanishing non-tensorial polarization modes in the Ricci-flat spacetime. It is consistent with the well-known fact that the Einstein gravity supports only two tensorial modes for the plane-wave gravitational waves on the flat background because of the Ricci-flat condition. The plane-wave condition \eqref{planecondition} enables to write these six conditions for the $z$-propagation in terms of the NP scalars
\begin{align}
\Psi_{1}=\Phi_{01},~\Psi_{2}=\Phi_{11}+\Lambda,~\Psi_{3}=\Phi_{21}\,.
\label{zcondition}
\end{align}
Since $\Phi_{11}$ and $\Lambda$ are real, the second condition implies that $\Psi_{2}$ is real. Substitution of these relations into the polarization amplitudes in \eqref{6} expresses them in terms of the nine NP scalars, 
$\Psi_{0},\Psi_{1},\Psi_{2},\Psi_{3},\Psi_{4},\Phi_{00},\Phi_{02},\Phi_{22},\Lambda$. Since $\Psi_{0},\Psi_{1},\Psi_{3},\Psi_{4},\Phi_{02}$ are complex, the nine NP scalars mean the fourteen components of the Riemann curvature tensor, which says that the NP scalars are inconvenient to describe the polarization amplitudes $p_{n}$s for the non-null geodesic.
In each $p_{n}$, the contributions are divided into two, the term which survives under the null condition and the terms in the square bracket, which vanish for null propagation,
\begin{align}
p_{1}^{(l)}
&=6\Psi_{2}-\left[ 2(\Psi_{2}+2\Lambda) \right]
\, ,
\nonumber\\
p_{2}^{(x)}
&={\rm 2Re}(\Psi_{3})-\left[2 {\rm Re}(\Psi_{1}) \right]
\, ,
\nonumber\\
p_{3}^{(y)}
&=-2{\rm Im}(\Psi_{3})-\left[2 {\rm Im}(\Psi_{1}) \right]
\, ,
\nonumber\\
p_{4}^{(+)}
&={\rm Re}(\Psi_{4})+\left[ {\rm Re}(\Psi_{0})-2{\rm Re}(\Phi_{02}) \right]
\, ,
\nonumber\\
p_{5}^{(\times )}
&=-\frac{1}{2}{\rm Im}(\Psi_{4})+\left[\frac{1}{2} {\rm Im}(\Psi_{0})-{\rm Im}(\Phi_{02}) \right]
\, ,
\nonumber\\
p_{6}^{(b)}
&=\Phi_{22}-\left[ 2(\Psi_{2}+2\Lambda)- \Phi_{00} \right]
\, .
\label{Ep}
\end{align}
It is easily checked that the null condition in \eqref{NPnull_relation} makes the deviation factors in the square brackets vanish.
In the scalar longitudinal ($p_{1}^{(l)}$) and breathing ($p_{6}^{(b)}$) modes, the common factor $\Psi_{2}+2\Lambda$ contributes to the deviation and $p_{6}^{(b)}$ has additional deviation by another NP scalar $\Phi_{00}$ of spin-weight $0$. The Weyl-NP scalars, $\Psi_{1}$ and $\Psi_{3}$ of spin-weights $\pm1$, are mixed in the vector-$x$ $(p_{2}^{(x)})$ and -$y$ $(p_{3}^{(y)})$ modes. The tensor component $\Psi_{4}$ is also mixed with the other scalars of spin-weights $\pm2$, $\Psi_{0}$, $\Phi_{02}$, and $\Phi_{20}(=\Phi_{02}^{*})$ in the plus $(p_{4}^{(+)})$ and cross $(p_{5}^{(\times)})$ polarization modes.
Consequently, the driving-force matrix \eqref{DFmatrix} for the plane-wave propagation becomes
\begin{align}
S_{{\textrm{\tiny plane}}}=
\begin{pmatrix}
\makecell{ \frac{1}{2}\{{\rm Re}(\Psi_{4})+\Phi_{22}~~~~~~~~~~~~~~~\\+[ {\rm Re}(\Psi_{0})-2{\rm Re}(\Phi_{02})~~~~\\- 2(\Psi_{2}+2\Lambda)+ \Phi_{00}]\} } & \makecell{ -\frac{1}{2}\{{\rm Im}(\Psi_{4})~~~~~~~~~~~~~~~~~~~~~~~\\-\left[ {\rm Im}(\Psi_{0})-2{\rm Im}(\Phi_{02}) \right]\} } & \makecell{ 2\{{\rm Re}(\Psi_{3})~~~~~~~~\\-\left[ {\rm Re}(\Psi_{1}) \right]\} } \\
\makecell{ -\frac{1}{2}\{{\rm Im}(\Psi_{4})~~~~~~~~~~~~~~~~~~~~\\
-\left[ {\rm Im}(\Psi_{0})-2{\rm Im}(\Phi_{02}) \right]\} } & \makecell{ -\frac{1}{2}\{{\rm Re}(\Psi_{4})-\Phi_{22}~~~~~~~~~~~~~~~\\+[ {\rm Re}(\Psi_{0})-2{\rm Re}(\Phi_{02})~~~~\\+ 2(\Psi_{2}+2\Lambda)- \Phi_{00}]\} }  & \makecell{ -2\{{\rm Im}(\Psi_{3})~~~~~~~~~\\+[ {\rm Im}(\Psi_{1}) ]\} } \\
\makecell{ 2\{{\rm Re}(\Psi_{3})~~~~~~~~\\-\left[ {\rm Re}(\Psi_{1}) \right]\} } & \makecell{ -2\{{\rm Im}(\Psi_{3})~~~~~~~~~~\\+\left[ {\rm Im}(\Psi_{1}) \right]\} } & \makecell{\!\! 6\{\Psi_{2}~~~~~~~~~~~~~~~~~~~\\-\left[ \frac{1}{3}(\Psi_{2}+2\Lambda) \right]\}\!\! } \\
\end{pmatrix}.
\label{S_plane}
\end{align}

Note that the terms in the square brackets in \eqref{Ep} are generally non-vanishing and it means that there are two sources of the deviation factors for non-null propagation of gravitational waves: One comes from the NP-null scalars in the first terms of \eqref{Ep} and the other comes from the other NP scalars in the square brackets of \eqref{Ep}. Therefore, computation and analysis of the polarization amplitudes for the non-null geodesic by using the NP-null scalars 
\eqref{pnbar}~\cite{Bessada:2009qw,dePaula:2004bc,Rizwana:2016qdq,Abedi:2017jqx,Alves:2010ms} are incorrect as far as the terms in the square bracket are non-vanishing. Thus the correction factors of deviation in the square brackets of \eqref{Ep} and/or \eqref{S_plane} should be taken into account in order to achieve the correct exact polarization amplitude for non-null propagation of the gravitational waves. Furthermore, for the non-null propagation of gravitational waves, $\Psi_{2}$ is mixed in the breathing mode $p_{6}^{(b)}$ in the last line of \eqref{Ep}, which implies that vanishing $\Phi_{22}$ does not mean vanishing breathing mode, $p_{6}^{(b)}=-2\Psi_{2}$, even when $\Lambda=0=\Phi_{00}$.

\subsection{Non-null propagation of gravitational waves in terms of metric perturbations}\label{Sec3.3}

In this subsection, we read the exact polarization amplitude $p_{n}$ from the driving-force matrix without relying on NP scalars. By taking into account the weak field assumption, the Riemann tensor is linearized as,
\begin{align}
R_{\mu\nu\rho\sigma}^{(1)}=-2\partial_{[\mu}\partial_{|[\rho}h_{\sigma]|\nu]}\,,
\end{align}
where the superscript ${}^{(1)}$ denotes the order in $h$. Then the polarization amplitudes $p_{n}$s are described in terms of the metric perturbation, 
\begin{align}
p_{1}^{({\textrm{l}})}\,\,&\approx  R_{tztz}^{(1)}~~~~~~~~~~~= -\frac{1}{2}(\partial_{t}^{2}h_{zz}-2\partial_{t}\partial_{z}h_{tz}+\partial_{z}^{2}h_{tt})\,,\nonumber\\
p_{2}^{(x)}\,&\approx R_{tztx}^{(1)}~~~~~~~~~~\,= -\frac{1}{2}(\partial_{t}^{2}h_{xz}-\partial_{t}\partial_{z}h_{tx})\,,\nonumber\\
p_{3}^{(y)}\,&\approx R_{tzty}^{(1)}~~~~~~~~~~\,= -\frac{1}{2}(\partial_{t}^{2}h_{yz}-\partial_{t}\partial_{z}h_{ty})\,,\nonumber\\
p_{4}^{(+)}&\approx  R_{txtx}^{(1)}-R_{tyty}^{(1)}= -\frac{1}{2}(\partial_{t}^{2}h_{xx}-\partial_{t}^{2}h_{yy})\,,\nonumber\\
p_{5}^{(\times )}&\approx R_{txty}^{(1)}~~~~~~~~~~\,= -\frac{1}{2}\partial_{t}^{2}h_{xy}\,,\nonumber\\
p_{6}^{(b)}\,&\approx R_{txtx}^{(1)}+R_{tyty}^{(1)}= -\frac{1}{2}(\partial_{t}^{2}h_{xx}+\partial_{t}^{2}h_{yy})\,.
\label{pi}
\end{align}
Since all the ten components of the metric perturbation $h_{\mu\nu}$ appear in the right-hand sides of \eqref{pi}, four redundant degrees should be removed by the gauge fixing. In the subsequent two subsubsections, we proceed the discussion under the Lorentz gauge condition and the Newtonian gauge condition.

\subsubsection{Lorentz gauge condition}

The production  and propagation of gravitational waves from various dynamical massive systems are calculated under the Lorentz gauge condition $\partial_{\mu}\bar{h}^{\mu\nu}\equiv \partial_{\mu}(h^{\mu\nu}-\frac{1}{2}\eta^{\mu\nu}h^{\lambda}_{~\lambda})=0$ in the Einstein gravity. Beyond it, this gauge is often used to describe the wave-like solutions. Four components of the Lorentz gauge condition are
\begin{align}
&\partial_{t}h_{tz}-\partial_{z}h_{xz}=0\,,\nonumber\\
&\partial_{t}h_{ty}-\partial_{z}h_{yz}=0\,,\nonumber\\
&(\partial_{t}^{2}-\partial_{z}^{2})h_{tz}=\partial_{t}\partial_{z}(h_{xx}+h_{yy})\,,\nonumber\\
&(\partial_{t}^{2}-\partial_{z}^{2})=-(\partial_{t}^{2}+\partial_{z}^{2})(h_{xx}+h_{yy})\,.
\label{Lg}
\end{align}
Removing the four time components, $h_{tt}$, $h_{tx}$, $h_{ty}$, $h_{tz}$, by application of the gauge fixing condition in \eqref{Lg}, we have a set of non-local expressions for the polarization amplitudes, 
\begin{align}
p_{1}^{(l)}\,\,&=-  \frac{1}{2}\left[\partial_{z}^{2}(h_{xx}+h_{yy})+\left( \partial_{t}^{2}-\partial_{z}^{2} \right)h_{zz}\right]\,,\nonumber\\
p_{2}^{(x)}\,&=-  \frac{1}{2} \left(\partial_{t}^{2} -\partial_{z}^{2} \right) h_{xz}\,,\nonumber\\
p_{3}^{(y)}\,&=- \frac{1}{2} \left( \partial_{t}^{2}-\partial_{z}^{2} \right) h_{yz},\nonumber\\
p_{4}^{(+)}&=- \frac{1}{2}\partial_{t}^{2}\left( h_{xx}-h_{yy} \right)\,,\nonumber\\
p_{5}^{(\times )}&=-  \frac{1}{2}\partial_{t}^{2}h_{xy},\nonumber\\
p_{6}^{(b)}\,&=- \frac{1}{2}\partial_{t}^{2}(h_{xx}+h_{yy})\, .
\label{lp}
\end{align}So far all the expressions in \eqref{lp} are still linear in metric perturbation and the would-be dynamical equation for $h_{\mu\nu}$ approximated in weak gravity limit is naturally expected to be a linear wave equation which supports the monochromatic wave solution of the form 
\begin{align}
h_{\mu\nu}=C_{\mu\nu}e^{-i\omega t+ikz}
\, ,
\label{mw}
\end{align}
where $\omega$ is the frequency and $k$ is the wave number. The linearity of the assumed wave equation guarantees that the spacetime-independent coefficients $C_{ij}$ are also independent of the frequency $\omega$ and the wave number $k$. Substitution of the monochromatic wave \eqref{mw}
into the gauge fixing condition \eqref{lp} leads to 
\begin{align}
&h_{tx}=-\frac{k}{\omega}h_{xz},~~h_{ty}=-\frac{k}{\omega}h_{yz},\nonumber\\
&h_{tz}=\frac{ \omega k}{\omega^{2}-k^{2}}(h_{xx}+h_{yy}),\nonumber\\
&h_{tt}=-h_{zz}-\frac{\omega^{2}+k^{2}}{\omega^{2}-k^{2}}(h_{xx}+h_{yy})\,,
\label{fg}
\end{align}
which tell us that the other four coefficients $C_{tt}$, $C_{tx}$, $C_{ty}$, $C_{tz}$ depend on the frequency $\omega$ and the wave  number $k$. Then the six polarization amplitudes $p_{n}$s are expressed in terms of six spatial components of the metric fluctuation. 
\begin{align}
p_{1}^{(l)}\,\,&=  \frac{1}{2}\left[k^{2}(h_{xx}+h_{yy})+\left( \omega^{2}-k^{2} \right)h_{zz}\right]\,,\nonumber\\
p_{2}^{(x)}\,&=  \frac{1}{2} \left( \omega^{2} -k^{2} \right) h_{xz}\,,\nonumber\\
p_{3}^{(y)}\,&= \frac{1}{2} \left( \omega^{2}-k^{2} \right) h_{yz},\nonumber\\
p_{4}^{(+)}&= \frac{1}{2}\omega^{2}\left( h_{xx}-h_{yy} \right)\,,\nonumber\\
p_{5}^{(\times )}&=  \frac{1}{2}\omega^{2}h_{xy},\nonumber\\
p_{6}^{(b)}\,&= \frac{1}{2}\omega^{2}(h_{xx}+h_{yy})\, .
\label{pL}
\end{align}
If the limit of the Einstein gravity is naively taken, the dispersion relation becomes $\omega^{2}=k^{2}$ and the four modes, $p_{1}^{(l)},~p_{4}^{(+)},~p_{5}^{(\times )},~p_{6}^{(b)}$ seem to be non-vanishing in \eqref{pL}, which is inconsistent with the fact that only the two tensor modes, $p_{4}^{(+)},~p_{5}^{(\times )}$, should survive. To reproduce correctly these physical modes, the transverse-traceless condition, $\partial_{\mu}h^{\mu}_{~\nu}=0$ and $h^{\mu}_{~\mu}=0$, should be additionally assigned.

It would be convenient to avoid this cumbersome assignment of an additional condition and to obtain the two tensor modes in the limit of the Einstein gravity. A specific way is to remove $h_{xx}$ from the physical components and  to include $h_{tt}$ as a physical mode. The corresponding monochromatic wave solution \eqref{mw} allows the new assumption on $C_{\mu\nu}$, i.e. \begin{align}\label{Ctt}
C_{tt},~C_{yy},~C_{zz},~C_{xy},~C_{yz},~C_{zx}
\end{align}
are independent of the frequency $\omega$ and the wave number $k$. On the other hand, the four gauge conditions in \eqref{fg} force the remaining four $C_{xx},~C_{tx},~C_{ty},~C_{tz}$ to be functions of $\omega$ and $k$, 
\begin{align}\label{gauge2}
&h_{tx}=-\frac{k}{\omega}h_{xz},~~h_{ty}=-\frac{k}{\omega}h_{yz},\nonumber\\
&h_{tz}=-\frac{ \omega k}{\omega^{2}+k^{2}}(h_{tt}+h_{zz}),\nonumber\\
&h_{xx}=-h_{yy}-\frac{\omega^{2}-k^{2}}{\omega^{2}+k^{2}}(h_{tt}+h_{zz})\,.
\end{align}
Thus, the gauge fixing condition \eqref{fg} reexpresses the polarization amplitudes $p_{n}$s as
\begin{align}
p_{1}^{(l)}\,\,&=  \frac{1}{2} \left( \frac{ \omega^{2}-k^{2} }{\omega^{2}+k^{2}} \right)\omega^{2}(h_{tt}+h_{zz})-\frac{1}{2}\left( \omega^{2}-k^{2} \right) h_{tt}\,,\nonumber\\
p_{2}^{(x)}\,&=  \frac{1}{2}\left( \omega^{2}-k^{2}\right) h_{xz}\,,\nonumber\\
p_{3}^{(y)}\,&=  \frac{1}{2}\left( \omega^{2}-k^{2}\right) h_{yz}\, , \nonumber\\
p_{4}^{(+)}&= -\frac{1}{2} \left( \frac{ \omega^{2}-k^{2} }{\omega^{2}+k^{2}} \right)\omega^{2}(h_{tt}+h_{zz}) -\omega^{2} h_{yy}\, , \nonumber\\
p_{5}^{(\times )}&=  \frac{1}{2} \omega^{2}h_{xy}\,,\nonumber\\
p_{6}^{(b)}\,&=-\frac{1}{2} \left( \frac{ \omega^{2}-k^{2} }{\omega^{2}+k^{2}} \right)\omega^{2}(h_{tt}+h_{zz}) \,. \label{p16complete}
\end{align}
The deviation from null geodesic appears through the separate term in every mode controlled by the non-vanishing common factor $\omega^{2}-k^{2}$ in the five polarization amplitudes $p_{1}^{(l)},~p_{2}^{(x)}$, $p_{3}^{(y)}$, $p_{4}^{(+)},~p_{6}^{(b)}$. Thus survival of the only two tensor modes in the limit of the null geodesic is automatically reproduced without any further condition by applying the dispersion relation $\omega=k$ which makes the common factor vanish, $\omega^{2}-k^{2}=0$. The magnitude of this additional effect is quantitatively determined by the specific form of the dispersion relation, $\omega=\omega(k)$. 
Accordingly, the NP-null scalars under the same gauge fixing condition \eqref{fg} are
\begin{align}
\Psi_{2}&=-\frac{1}{24}\left( \frac{\omega^{2}-k^{2}}{\omega^{2}+k^{2}} \right)[(3k^{2}-\omega^{2})h_{tt}+(k^{2}-3\omega^{2})h_{zz}]\,,\nonumber\\
\Psi_{3}&=\frac{1}{8}\frac{(\omega-k)(\omega+k)^{2}}{\omega}(h_{xz}-ih_{yz})\,,\nonumber\\
\Psi_{4}&=-\frac{1}{8}\frac{(\omega-k)(\omega+k)^{3}}{\omega^{2}+k^{2}}(h_{tt}+h_{zz})-\frac{1}{4 }(\omega+k)^{2}(h_{yy}+ih_{xy})\,,\nonumber\\
\Phi_{22}&=-\frac{1}{8}\frac{(\omega-k)(\omega+k)^{3}}{\omega^{2}+k^{2}}(h_{tt}+h_{zz})\,.
\label{corrected_Paula}
\end{align}

\subsubsection{Newtonian gauge condition}

When the general metric perturbations are decomposed in the basis of the representations of the spatial rotation, all the sixteen components have
\begin{align}\label{nwg}
&\delta g_{00}=-2A\,,\nonumber\\
&\delta g_{0i}=-\partial_{i}B-B_{i}\,,\nonumber\\
&\delta g_{ij}=-2\delta_{ij}D+2\left( \partial_{i}\partial_{j}-\frac{\delta_{ij}}{3}\partial^{k}\partial_{k} \right)E+2\partial_{(i}E_{j)}+h_{ij}\,,
\end{align}
where the symmetric property is recovered by the following six rotations, $\partial^{i}B_{i}=0$, $\partial^{i}E_{i}=0$, $\partial^{i}h_{ij}=0$, and $h^{i}_{~i}=0$. Then the ten modes are decoupled at the linear level of this decomposition. In the representation of the spatial rotation about the specific $\hat{z}$ axis, the metric perturbation $h_{\mu\nu}$ takes the matrix form as
\begin{align}
h_{\mu\nu}=\begin{pmatrix}\displaystyle 
-2A & -B_{x} & -B_{y} & -B_{,z} \\
-B_{x} & \displaystyle -2D-\frac{2}{3}E_{,zz}+h_{+} & \displaystyle h_{\times } & E_{x,z}\\
-B_{y} & \displaystyle h_{\times } & \displaystyle -2D-\frac{2}{3}E_{,zz}-h_{+} & E_{y,z}\\
-B_{,z} & E_{x,z} & E_{y,z} & \displaystyle -2D+\frac{4}{3}E_{,zz}\\
\end{pmatrix} \,\label{p6SVT} .
\end{align} 
Insertion of this into the six polarization amplitudes $p_{n}$s~\eqref{pi} leads to
\begin{align}
p_{1}^{(l)}\,\,&=\partial_{t}^{2}  D-\frac{2}{3}\partial_{t}^{2}(E_{,zz})-\partial_{t}\partial_{z}(B_{,z})+\partial_{z}^{2}A\,,\nonumber\\
p_{2}^{(x)}\,&=  -\frac{1}{2}[\partial_{t}^{2}(E_{x,z})+\partial_{t}\partial_{z}B_{x}]\,,\nonumber\\
p_{3}^{(y)}\,&=  -\frac{1}{2}[\partial_{t}^{2}(E_{y,z})+\partial_{t}\partial_{z}B_{y}]\,,\nonumber\\
p_{4}^{(+)}&= -\partial_{t}^{2}h_{+}\,,\nonumber\\
p_{5}^{(\times )}&=  -\frac{1}{2}\partial_{t}^{2}h_{\times }\,,\nonumber\\
p_{6}^{(b)}\,&= 2\partial_{t}^{2}D+\frac{2}{3}\partial_{t}^{2}(E_{,zz})\,.\label{p16}
\end{align}
An appropriate gauge fixing condition for this decomposition is the conformal Newtonian gauge, $B_{x}=B_{y}=B=E=0$, which results in $h_{tx} = h_{ty} = h_{tz} = 0$ and $h_{xx}+h_{yy}=2h_{zz}$. Under this gauge fixing condition, the above six polarization amplitudes $p_{n}$s become 
\begin{align} 
p_{1}^{(l)}\,\,&=\partial_{t}^{2}  D+\partial_{z}^{2}A\,,\nonumber\\
p_{2}^{(x)}\,&=  -\frac{1}{2}\partial_{t}^{2}(E_{x,z})\,,\nonumber\\
p_{3}^{(y)}\,&=  -\frac{1}{2}\partial_{t}^{2}(E_{y,z})\,,\nonumber\\
p_{4}^{(+)}&= -\partial_{t}^{2}h_{+}\,,\nonumber\\
p_{5}^{(\times )}&=  -\frac{1}{2}\partial_{t}^{2}h_{\times }\,,\nonumber\\
p_{6}^{(b)}\,&= 2\partial_{t}^{2}D\,. \label{p6Newtonian}
\end{align}
For the monochromatic waves, \eqref{p6Newtonian} give
\begin{align} 
p_{1}^{(l)}\,\,&=-\omega^{2}  D-k^{2}A\,,\nonumber\\
p_{2}^{(x)}\,&=  \frac{1}{2}\omega^{2}(E_{x,z})\,,\nonumber\\
p_{3}^{(y)}\,&=  \frac{1}{2}\omega^{2}(E_{y,z})\,,\nonumber\\
p_{4}^{(+)}&= \omega^{2}h_{+}\,,\nonumber\\
p_{5}^{(\times )}&=  \frac{1}{2}\omega^{2}h_{\times }\,,\nonumber\\
p_{6}^{(b)}\,&= -2\omega^{2}D\,. \label{p6Newtonian_w}
\end{align}

\subsection{Response function}\label{Sec3.4}

In the detectors of the gravitational waves,  the phase difference between the light
signals traveling in both arms of the interferometer is given by 
\begin{align}
\Delta \Phi=2\pi\nu(2L_{1}-2L_{2})\equiv 2\pi \nu L_{0}S(t)\,,
\end{align}
where $\nu$ is the frequency of the laser light, $L_{0} $ is the length of the unperturbed interferometer arm, $L_{1}$ and $L_{2}$ are the perturbed lengths of two arms, and $S(t)$ is the detector's response 
function~\cite{Will:2014kxa,Isi:2015cva}. The response function $S(t)$ is written in terms of the theoretically-obtained polarization amplitude $p_{n}$ multiplied by the normalization coefficients $a_{n}$ of the basis polarization matrices in \eqref{basis_matrix} and the angular pattern function $F_{n}$,
\begin{align} 
S(t) =\sum_{n=1}^{6}2\tilde{p}_{n}a_{n}F_{n}
\, ,
\label{Sr}
\end{align}
where $p_{n}\equiv -\ddot{\tilde{p}}_{n}$. The angular pattern functions $F_{n}$'s have five different components as in 
Ref.~\cite{Will:2014kxa,Isi:2015cva}, 
\begin{align} 
F_{b} &= -\frac{1}{2} \sin^{2} \theta \cos 2 \phi = - F_{l} \label{Fb0} \, , \\
F_{x} &= - \sin \theta \left( \cos \theta \cos 2 \phi \cos \psi - \sin 2 \phi \sin \psi \right) \label{Fx0} \, , \\
F_{y} &= - \sin \theta \left( \cos \theta \cos 2 \phi \sin \psi + \sin 2 \phi \cos \psi \right) \label{Fy0} \, , \\
F_{+} &= \frac{1}{2} \left( 1 + \cos^{2} \theta \right) \cos 2 \phi \cos 2 \psi - \cos \theta \sin 2 \phi \sin 2 \psi \label{Fpl0} \, , \\
F_{\times} &= \frac{1}{2} \left( 1 + \cos^{2} \theta \right) \cos 2 \phi \sin 2 \psi - \cos \theta \sin 2 \phi \cos 2 \psi  \label{Fcross0} \, .
\end{align} As far as the gravitational wave along the light trajectory is weak, the response function is generally given by a superposition of the contributions of monochromatic gravitational waves. For each monochromatic wave of frequency $\omega$, it satisfies $\tilde{p}_{n}=-\frac{1}{\omega^{2}}p_{n}$, and thus the response function \eqref{Sr} becomes
\begin{align} 
S_{n}(t)=\sum_{n=1}^{6}2\frac{p_{n}}{\omega^{2}}a_{n}F_{n}
\, . 
\label{St0} 
\end{align}

Note that the value of each $p_{n}a_{n}$ is independent of the choice of the basis matrix $a_{n}$ \eqref{basis_matrix} and each response function $S_{n}$,
\begin{align}
S_{n}=2\tilde{p}_{n}a_{n}F_{n}=-\frac{2}{\partial_{t}^{2}}p_{n}a_{n}F_{n}
,
\end{align}
is gauge-invariant. We read six response functions in terms of metric components in the non-local expression,

\begin{align}
&S^{(l)}=\frac{1}{\partial_{t}^{2}}(\partial_{t}^{2}h_{zz}-2\partial_{t}\partial_{z}h_{tz}+\partial_{z}^{2}h_{tt}) F_{l}\,,\nonumber\\
&S^{(x)}=\frac{1}{\partial_{t}^{2}}(\partial_{t}^{2}h_{xz}-\partial_{t}\partial_{z}h_{tx})  F_{x}\,,\nonumber\\
&S^{(y)}=\frac{1}{\partial_{t}^{2}} (\partial_{t}^{2}h_{yz}-\partial_{t}\partial_{z}h_{ty})  F_{y}\,,\nonumber\\
&S^{(+)}=\frac{1}{2\partial_{t}^{2}} \partial_{t}^{2}(h_{xx}-h_{yy}) F_{+}\,,\nonumber\\
&S^{(\times )}=\frac{1}{\partial_{t}^{2}}  \partial_{t}^{2}h_{xy}  F_{\times }\,,\nonumber\\
&S^{(b)}=\frac{1}{2\partial_{t}^{2}}  \partial_{t}^{2}(h_{xx}+h_{yy})  F_{b}\,.
\end{align}
As in \eqref{Fb0} the angular pattern functions of the longitudinal mode and the breathing mode are the same, $F_{b}=-F_{l}$, and the breathing and the longitudinal pattern functions are degenerated. Thus no array of laser interferometers can measure their two modes separately~\cite{Will:2014kxa}.
In addition to the four pattern functions $S^{(x)},S^{(y)},S^{(+)},S^{(\times)}$, the single response function given by the sum of longitudinal and breathing modes,
\begin{align}
S^{(l+b)}
=\frac{1}{\partial_{t}^{2}}\left\{ \partial_{t}^{2}\left[ h_{zz}-\frac{1}{2}(h_{xx}+h_{yy}) \right] -2\partial_{t}\partial_{z}h_{tz}+\partial_{z}^{2}h_{tt}\right\}F_{l}
\, ,
\end{align}
is taken into account.
When a monochromatic wave \eqref{mw} is assumed, the five response functions become
\begin{align}
&S^{(x)}=(h_{xz}+\frac{k}{\omega} h_{tx})  F_{x}\,,\nonumber\\
&S^{(y)}= (h_{yz}+\frac{k}{\omega}h_{ty})  F_{y}\,,\nonumber\\
&S^{(+)}=\frac{1}{2}(h_{xx}-h_{yy}) F_{+}\,,\nonumber\\
&S^{(\times )}=  h_{xy} F_{\times }\,,\nonumber
\end{align}
and
\begin{align}
S^{(l+b)}&=\left\{ \left[ h_{zz}-\frac{1}{2}(h_{xx}+h_{yy}) \right] +2\frac{k}{\omega} h_{tz}+\frac{k^{2}}{\omega^{2}}h_{tt}\right\}F_{l}\,.
\label{5S}
\end{align} 

\subsubsection{Lorentz gauge condition}

When the Lorentz gauge condition \eqref{Lg} is chosen, the polarization amplitudes $p_{n}$s are already obtained for a monochromatic wave in
\eqref{pL} and thus the five components of the response function are obtained. From \eqref{5S}, the response function for breathing and longitudinal modes becomes
\begin{align}\label{Sbl}
S^{(b+l)}
&= \left\{\frac{\left( \omega^{2}-k^{2} \right)}{\omega^{2}}[h_{zz}-(h_{xx}+h_{yy})] +   \frac{1}{2}(h_{xx}+h_{yy})  \right\}F_{l}
\, ,
\nonumber
\end{align}
and the other four modes are
\begin{align}
&S^{(x)}
=\frac{\omega^{2}-k^{2}}{\omega^2}  h_{xz}F_{x}
\, ,\nonumber\\
&S^{(y)}
=\frac{\omega^{2}-k^{2}}{\omega^2}  h_{yz}F_{y}
\, ,\nonumber\\
&S^{(+)}
=\frac{1}{2}\left( h_{xx}-h_{yy} \right) F_{+}
\, ,\nonumber\\
&S^{(\times )}
=h_{xy}F_{\times }
\, .
\end{align}
As we already discussed, the response function for the breathing and longitudinal modes \eqref{Sbl} does not vanish even in the null limit of $\omega^{2}=k^{2}$ under the consideration of constant $h_{ij}$. Thus, under the Lorentz gauge, a convenient choice is to set $C_{tt}$ a constant as in \eqref{gauge2} instead of $C_{xx}$. Note again that the six amplitudes of the gravitational wave, $C_{tt}$, $C_{yy}$, $C_{zz}$, $C_{xy}$, $C_{yz}$, $C_{yx}$, do not depend on the frequency $\omega$ and the wave number $k$, that makes the detection of the polarization tractable. By using the expression \eqref{p16complete}, the response function for the breathing and longitudinal modes becomes
\begin{align}
S^{(b+l)}
&= \frac{(\omega^{2}-k^{2})}{2\omega^{2}(\omega^{2}+k^{2})}\left[ (\omega^{2}-2k^{2})h_{tt}+3\omega^{2}h_{zz} \right]F_{l}\,,
\nonumber
\end{align}
and those for the other four modes are
\begin{align}
&S^{(x)}
=\frac{\omega^{2}-k^{2}}{\omega^2}  h_{xz}F_{x}
\, ,\nonumber\\
&S^{(y)}
=\frac{\omega^{2}-k^{2}}{\omega^2}  h_{yz}F_{y}
\, ,\nonumber\\
&S^{(+)}
=\left[ -h_{yy}-\frac{1}{2}\frac{\omega^{2}-k^{2}}{\omega^{2}+k^{2}}(h_{tt}+h_{zz}) \right] F_{+}
\, ,\nonumber\\
&S^{(\times )}
=h_{xy}F_{\times }
\, .
\label{Sbo}
\end{align}
As expected, there is an overall $(\omega^{2}-k^{2})$ factor in $S^{(b+l)}$, $S^{(x )}$, and $S^{(y)}$, which lets those vanish continuously in the null limit.

\subsubsection{Newtonian gauge condition}

Similar to the Lorentz gauge condition, the response function for breathing and longitudinal modes is given under the Newtonian gauge condition \eqref{nwg} as
\begin{align}
S^{(b+l)}
=\frac{k^{2}}{\omega^{2}}h_{tt} F_{l}
\, , \label{Sbl}
\end{align}
and the other four amplitudes are
\begin{align}
&S^{(x)}
=h_{xz}  F_{x}
\, ,\nonumber\\
&S^{(y)}
= h_{yz}  F_{y}
\, ,\nonumber\\
&S^{(+)}
=h_{+}  F_{+}
\, ,\nonumber\\
&S^{(\times )}
= h_{xy} F_{\times }
\, .
\label{Sbx}
\end{align}
Since all five amplitudes of the monochromatic gravitational wave are constants, the four response functions, $S^{(x)},S^{(y)},S^{(+)},S^{(\times)}$, involve no dependence on the frequency $\omega$ and the wave number $k$, however  that of the breathing and longitudinal modes $S^{(b+l)}$ only  depends on.


\section{Model calculation}\label{Sec4}

The discussion up to the section \ref{Sec3} has been made without assuming a specific form of the wave equation, equivalently the form of the action for gravity, and thus the polarization amplitudes $p_{n}$s \eqref{pL} can be applicable to any null or non-null propagation of gravitational wave from arbitrary metric-compatible gravity theories. From now on, various gravity models are considered and examined, including alternative models of gravity. To investigate the behavior of the six polarization amplitudes, we already constructed the formalism and thus need only to specify the dispersion relation, $\omega=\omega(k)$, according to the model of interest. We introduce various gravity models away from the Einstein gravity and examine the polarization amplitudes in what follows. 

\subsection{Mass effect}\label{Sec4.1}

Even when the wave equation does not involve higher-derivative terms, the relativistic relation between energy and momentum does not prohibit the mass term
\begin{align} 
E^2 = p^2 + m_{{\textrm{g}}}^2 
\label{deBDR1} 
\end{align}
whose dispersion relation is $\omega=\pm\sqrt{m_{{\textrm{g}}}^{2}+k^{2}}$.
Since the general covariance protects the introduction of a mass term in metric-compatible gravity theories and the Pauli-Fierz type mass term for spin-2 field is ruled out, a possibility to introduce the mass term with $g_{\mu\nu}$ is to employ the bi-metric theory in which both the background metric $g_{\mu\nu}^{0}$ and the metric for the gravitational field $(g-g_{0})_{\mu\nu}$ are tensor quantities \cite{Visser:1997hd}.

\begin{figure}[H]
\centering
\includegraphics[scale=0.6]{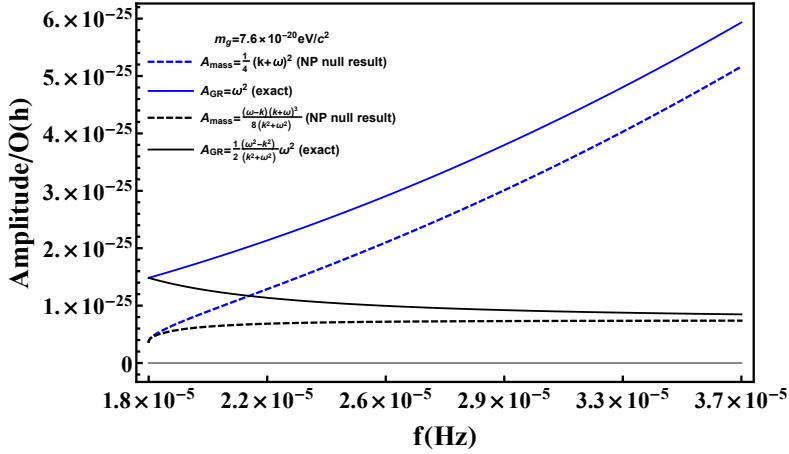}
\caption{The behavior of the exact (solid lines) and approximate (dashed lines) polarization amplitudes are compared by choosing a mass parameter $m_{{\textrm{g}}}=7.6\times 10^{-20}{\textrm{eV}}/{\textrm{c}}^{2}$ of the dispersion relation $\omega=\sqrt{m_{{\textrm{g}}}^{2}+k^{2}}$. The mode amplitude appearing for the time-like geodesic and that for the null geodesic are also compared by the black and blue colored lines.}
\label{MassiveEff}
\end{figure}

The dispersion relation $\omega=\sqrt{m_{{\textrm{g}}}^{2}+k^{2}}$ can be used in weak gravity limit as far as the bi-metric theory is considered. In $p_{4}^{(+)}$\eqref{p16complete}, the second term  $ -\omega^{2} h_{yy} $ is the plus mode amplitude for the null geodesic and the first term $-\frac{1}{2} \left( \frac{ \omega^{2}-k^{2} }{\omega^{2}+k^{2}} \right)\omega^{2}(h_{tt}+h_{zz})$ appears for the time-like geodesic, which also coincides with the breathing mode amplitude $p_{6}^{(b)}$. These two exact amplitudes in $p_{4}^{(+)}$ \eqref{pn} are compared to the corresponding approximate amplitudes of ${\textrm{Re}}(\Psi_{4})$ \eqref{corrected_Paula} and the result is given in Fig.~\ref{MassiveEff}. The blue and black solid lines denote $\omega^{2}$ and $\frac{1}{2}\frac{\omega^{2}-k^{2}}{\omega^{2}+k^{2}}\omega^{2}$ in the exact amplitude, and the blue and black dashed lines do $\frac{1}{4}(\omega+k)^{2}$ and $\frac{1}{8}\frac{(\omega-k)(\omega+k)^{3}}{\omega^{2}+k^{2}}$ in the approximate result, respectively. The graphs show that behavior of the exact polarization amplitudes is different from that of the approximate polarization amplitudes obtained by NP-null scalars.

\begin{figure}[H]
\centering
\includegraphics[scale=0.6]{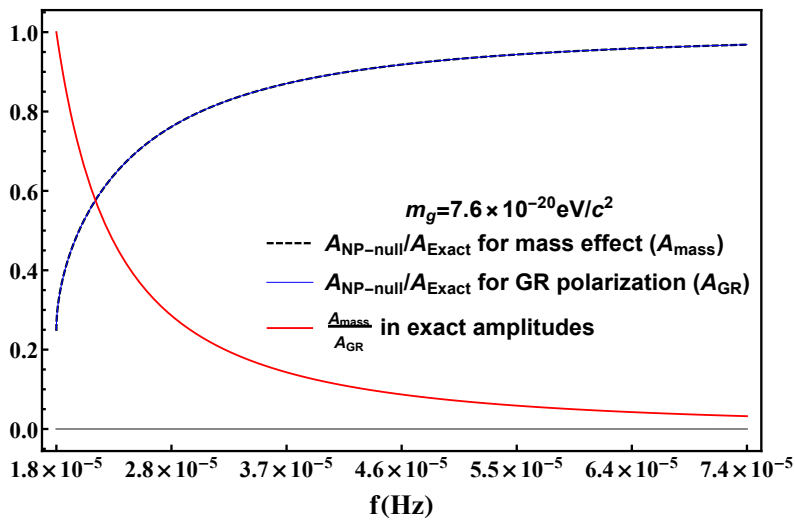}
\caption{The ratio between the approximate and exact mode amplitudes in the low-frequency region is shown for the weakest model-independent graviton mass bound. The more the mode amplitude appearing for the time-like geodesic is enhanced, the more the deviation of the approximate amplitude from the exact one grows.}
\label{ratio}
\end{figure}

Comparison of the two solid lines in Fig.~\ref{MassiveEff} show the followings. As easily expected, the effect due to the time-like geodesic becomes negligible in the high-frequency region. In the low-frequency region, the mode amplitude appearing for the time-like geodesic is magnified and becomes comparable to the mode amplitude for the null geodesic. The analogous conclusion was pointed out in the context of the approximate amplitudes obtained with the NP-null scalars \cite{dePaula:2004bc} as also given by the dashed lines in Fig.~\ref{MassiveEff}. The weakest bound of the graviton mass $m_{{\textrm{g}}}=7.6\times 10^{-20}{\textrm{eV/c}}^{2}$ is chosen in Fig.~\ref{MassiveEff} from the various model-independent mass bounds of the graviton which are  listed in Table \ref{mgtable}. The mode amplitude appearing for the time-like geodesic is significantly enhanced in the frequency region about $2\times 10^{-5}$ Hz. The frequency regions of maximum enhancement are summarized in Table \ref{Table_fc}, which are overlapped with the frequency domain of the future detector, such as the pulsar timing arrays with the range of $10^{-9}\sim 10^{-7}$ Hz \cite{daSilvaAlves:2011fp,Chamberlin:2011ev}. As far as the amplitudes for model-dependent (MD) mass bounds of the graviton are concerned, the frequency region for the maximum enhancement appears in an ultra-low frequency region which has been dealt in various inflation  models but is too low to be detected by the future detectors planned. As the detection level is lifted up with more accurate values, the suggested enhanced effect of the polarization modes due to the time-like geodesic may have more chance to be detected. In the very low-frequency region with the maximum enhancement, the deviation of the approximate result from the exact result also increases significantly, which is shown clearly in Fig.~\ref{ratio}. Therefore, the exact formalism based on the time-like geodesic will play an important role in order to compare the theoretical results with future observed data. 

\begin{table}[H]
\centering
\begin{tabular}{ccccc}\hline\hline
$\lambda_{{\textrm{g}}}$(km) & $m_{{\textrm{g}}}$(eV/c${}^{2}$) & Observation & Properteis & References \\\hline
$2.8\times 10^{12}$&$4.4\times 10^{-22}$ & solar system & static, MID & \cite{Talmadge:1988qz,Will:1997bb} \\\hline
$2.5\times 10^{13}$&$5.0\times 10^{-23}$ & supermassive black hole & static, MID & \cite{Brito:2013wya} \\\hline
$6.2\times 10^{19}$&$2.0\times 10^{-29}$ & galactic clusters & static, MD & \cite{Goldhaber:1974wg} \\
$9.1\times 10^{19}$&$1.37\times 10^{-29}$& galaxy cluster Abell 1689 & static, MD & \cite{Desai:2017dwg} \\ \hline
$1.8\times 10^{22}$&$6.9\times 10^{-29}$ & weak lensing & static, MD & \cite{Choudhury:2002pu}\\\hline
$1.63\times 10^{10}$&$7.6\times 10^{-20}$ & binary pulsars & dynamical, MID & \cite{Finn:2001qi} \\\hline
$1.0\times 10^{13}$&$1.2\times 10^{-22}$ & binary black holes & dynamical, MID & \cite{TheLIGOScientific:2016src}\\\hline
\end{tabular}
\caption{The lower bounds of Compton wavelength of the massive graviton, $\lambda_{{\textrm{g}}}$, and its corresponding upper bounds of the graviton mass, $m_{{\textrm{g}}}=h/\lambda_{{\textrm{g}}}c$, from different observations. MD and MID mean model-dependent and model-independent, respectively. For the details on the bounds, refer to \cite{deRham:2016nuf,Lee:2017dox}.}
\label{mgtable}
\end{table}
\begin{table}[H]
\centering
\begin{tabular}{ccc}\hline\hline
 $m_{{\textrm{g}}}$(eV/c${}^{2}$) & Observation & Frequency of massive effects (Hz)  \\\hline
$4.4\times 10^{-22}$ & solar system & $1.06\times 10^{-7}$  \\\hline
$5.0\times 10^{-23}$ & supermassive black holes & $1.21\times 10^{-8}$  \\\hline
$2.0\times 10^{-29}$ & galactic clusters & $4.84\times 10^{-15}$ \\\hline
$6.9\times 10^{-29}$ & weak lensing & $1.67\times 10^{-14}$ \\\hline
$7.6\times 10^{-20}$ & binary pulsars & $1.84\times 10^{-5}$  \\\hline
$1.2\times 10^{-22}$ & binary black holes & $2.90\times 10^{-8}$ \\\hline
\end{tabular}
\caption{The table shows the frequency regions where the massive effect becomes comparable to the massless one on the polarization amplitudes.}
\label{Table_fc}
\end{table}
\vspace{12pt}


\subsection{Modified gravity theories}\label{Sec4.2}

In this subsection, we consider the modified gravity theories with a generalized dispersion relation. As discussed in Ref~\cite{Mirshekari:2011yq,Kiyota:2015dla}, the generalized dispersion relation to cover almost all theories of interest takes
\begin{align} 
E^2 = p^2 c^2 + m_{{\textrm{g}}}^2 c^4 + A p^{\alpha} c^{\alpha} \, , 
\label{GDP}
\end{align}
where the two parameters, $A$ and $\alpha$, express signal of Lorentz violation.
The speed of graviton satisfying $E=\hbar\omega$ and $p=\hbar k$ is
\begin{align} 
\frac{v_{\textrm{g}}^2}{c^2} &\equiv \frac{1}{c^2} \left( \frac{d \omega}{dk} \right)^2 = 1 - \frac{4m_{{\textrm{g}}}^2 c^4 - 4 A p^{\alpha} c^{\alpha} (\alpha-1) - A^2 \alpha^2 p^{2(\alpha-1)} c^{2(\alpha-1)}   }{4 E^2} \, , \label{vgEDR} 
\end{align}
where the causality requires the numerator of the second term to be non-negative, $m_{\textrm{g}} c^2 \geq p^{\alpha-1} c^{\alpha-1} \sqrt{A (\alpha-1) p^2 c^2 + (\frac{A \alpha}{2} )^2} $.\\

An interesting case is $\alpha=1$ which is a non-local theory including $p=\sqrt{p_{x}^{2}+p_{y}^{2}+p_{z}^{2}}$. When $\alpha=1$, the numerator of the second term in \eqref{vgEDR} becomes a constant and thus comparison with the usual mass case, $\displaystyle \frac{v_{\textrm{g}}^2}{c^2} = 1 - \frac{m_{{\textrm{g}}}^2 c^4}{E^2}$, leads to the effective mass $m_{\textrm{g}}^{\textrm{eff}} = \sqrt{m_{\textrm{g}}^2 - A^{2}/(4 c^4)}$. An example of this effective mass $m^{\text{eff}}_{{\textrm{g}}}$ appeared in the $f(R)$-gravity theories \cite{Lee:2017dox}. A few higher power cases of higher derivatives, e.g. $\alpha=3,4$, are already discussed in Ref. \cite{Mirshekari:2011yq,Kiyota:2015dla}.\\

From now on, let us consider the extra-dimensions. In the model of an extra-dimension with $A=-\frac{\eta_{\text{ED}}}{E_{p}^2}$ and $\alpha= 4$, the generalized dispersion relation at low energy is given by $k^2 = \omega^2 + (\eta_{\text{ED}}/E_{p}^2) \omega^4 - m_{{\textrm{g}}}^2$~\cite{Sefiedgar:2010we}. Under the Newtonian gauge, the response function for breathing and longitudinal modes \eqref{Sbl} at a frequency $\omega_{1}$ is 
\begin{align} 
S^{(b+l)}[\omega_{1}] \equiv S^{(b+l)}_{1}= F_{l}  \left( 1+\frac{\eta_{\text{ED}}}{E_{p}^2} \omega_{1}^2 - m_{{\textrm{g}}}^2 \right) h_{tt} \label{SbliED} \, . 
\end{align}
Similarly, for the second frequency $\omega_{2}~(\omega_{2}\neq \omega_{1}),$
 the difference of the response functions is 
\begin{align} \label{SblimjED}
S^{(b+l)}_{1} - S^{(b+l)}_{2} = F_{l} \Biggl( \frac{\eta_{\text{ED}}}{E_{p}^2} + \frac{m_{{\textrm{g}}}^2}{\omega_{1}^2 \omega_{2}^2} \Biggr) \left( \omega_{1}^2 - \omega_{2}^2 \right) h_{tt} \, . 
\end{align}
Since the right-hand side of \eqref{SblimjED} involves the three unknown quantities, $\eta_{{\textrm{ED}}}/E_{p}^{2}$, $m_{{\textrm{g}}}^{2}$, and $h_{tt}$, we consider the third frequency $\omega_{3}$ different from $\omega_{1}$ and $\omega_{2}$ and then the expressions for $\eta_{{\textrm{ED}}}/E_{p}^2$ and $m_{{\textrm{g}}}^2$ are obtained only by the measured quantities, $S_{i}^{(b+l)}$ and $\omega_{i}$ $(i=1,2,3)$,
\begin{align} 
\frac{\eta_{\text{ED}}}{E_{p}^2} &=\omega _1^2 \omega _2^2 \omega_{2}^{3}\frac{\left(\omega _3^{-2}-\omega _2^{-2}\right) S_1^{(b+l)}+\left(\omega _2^{-2}-\omega _1^{-2}\right) S_3^{(b+l)}+\left(\omega _1^{-2}-\omega _3^{-2}\right) S_2^{(b+l)}}{\omega _1^2 \left(\omega _3^4-\omega _2^4\right) S_1^{(b+l)}+\omega _3^2\left(\omega _2^4-\omega _1^4\right)  S_3^{(b+l)}+\omega _2^2 \left(\omega _1^4-\omega _3^4\right) S_2^{(b+l)}}\,,  \nonumber\\
m_{{\textrm{g}}}^2 &=\omega _1^2 \omega _2^2 \omega _3^2\frac{ \left(\omega _3^2-\omega _2^2\right) S_1^{(b+l)}+\left(\omega _2^2-\omega _1^2\right) S_3^{(b+l)}+\left(\omega _1^2-\omega _3^2\right) S_2^{(b+l)}}{\omega _1^2 \left(\omega _3^4-\omega _2^4\right) S_1^{(b+l)}+\omega _3^2\left(\omega _2^4-\omega _1^4\right)  S_3^{(b+l)}+\omega _2^2 \left(\omega _1^4-\omega _3^4\right) S_2^{(b+l)}}\label{mg2htt} \, . 
\end{align}
For the other various modified gravity model, the results are also summarized in the table \eqref{tab-app1} \cite{Mirshekari:2011yq} with the generalized dispersion relations and the references. 
\begin{table}[H]
\centering
\label{tab-app1}
\begin{tabular}{ccccc}
\hline \hline 
Models     & $A$  & $\alpha$ & $m_{g}$  & References  
\\
\hline
Double Special Relativity & \multirow{2}{*}{$\frac{\eta_{\text{DSR}}}{E_{p}}$} & \multirow{2}{*}{3} & & \multirow{2}{*}{ 
\cite{AmelinoCamelia:2000ge,Magueijo:2001cr,AmelinoCamelia:2002wr,AmelinoCamelia:2010pd}} \\ 
Broken-Symmetry               &  &  &  &  \\
\cdashline{1-5}
Extra-Dimension                &  $-\frac{\eta_{\text{ED}}}{E_{p}^2}$ & 4 &   & \cite{Sefiedgar:2010we} \\
\cdashline{1-5}
Ho\v{r}ava-Lifshitz  &  $\frac{\kappa_{\text{HL}}^4 \mu_{\text{HL}}^2}{16}$ & 4 & 0 & \cite{Horava:2008ih,Horava:2009uw,Bogdanos:2009uj,Vacaru:2010rd,Blas:2011zd} \\
\cdashline{1-5}
Non-commutative Geometries & $2\frac{\eta_{\text{NCG}}}{E_{p}^2}$ & 4 &  & \cite{Garattini:2011kp,Garattini:2011hy}  \\
\hline
\end{tabular}
\caption{The generalized dispersion relations for various gravity models are displayed. Here, $E_{p}$ is the Planck energy scale, $\eta_{\text{DSR}}$  a dimensionless parameter given by the Lorentz invariance violating theories,  $\eta_{\text{ED}}$ a positive dimensionless parameter, $\kappa_{\text{HL}}$ and $\mu_{\text{HL}}$ constants of the Ho\v{r}ava-Lifshitz theory, and $\eta_{\text{NCG}}$ is a constant in the theory of non-commutative geometries.}
\end{table}


When the extended gravity theories involving propagating massive degrees are considered, the exact amplitude expressions \eqref{Ep}, \eqref{pi}, and \eqref{p16} can always be used to obtain the six mode polarizations of gravitational waves, (two scalars, two vectors, two tensors), irrespective of the form of their actions. Under the Newtonian gauge condition, non-vanishing components in the scalar-tensor gravity are $\delta \phi = A = -D$ and 
$h_{+}, h_{\times}$~\cite{Liang:2017ahj}. For other models, e.g., the Einstein-\AE ther theory, TeVeS, etc., the amplitudes are also obtained specifically~\cite{Gong:2018cgj}.

\section{Conclusion}\label{Conclusion}

We first extend the NP formalism to describe not only the null geodesic but also the time-like geodesic which is necessary for massive gravity theories. The amplitudes of the six polarization modes are exactly obtained in terms of the metric perturbation via the driving-force matrix \eqref{S_plane} under a few gauge fixing conditions: \eqref{lp} and \eqref{pL} for the Lorentz gauge, \eqref{p16complete} for our gauge choice, and \eqref{p6Newtonian} and \eqref{p6Newtonian_w} for the Newtonian gauge. For a given frequency, the corresponding distinctive five response functions are constructed in \eqref{Sbo}, \eqref{Sbl}, and \eqref{Sbx}, respectively. The method to distinguish the breathing and longitudinal mode via three frequency measurement is also proposed. The formulas throughout this work are applicable to every metric-compatible gravity theories. Though various theories have already been examined by use of the formalism valid only for the null geodesic, the theories including the non-null geodesic should be reexamined. Accordingly, the exact six polarization amplitudes are computed and analyzed in various extended gravity theories as examples. Since the stage of measuring gravitational waves moves to the search of the signal coming from a theory beyond the Einstein's GR, our construction of the general formalism seems timely and will become more important.

Our final comment is about the classification of extended gravity theories. In Ref~\cite{Eardley:1974nw}, the null condition was used to classify the extended gravity theories, the E(2) classification by using the little group of the polarization NP-null scalars. As explained in \cite{Eardley:1974nw}, the little group of the general Lorentz transformation for massless particles is given by the two-dimensional Euclidean group. In the case of the time-like propagation, the little group of the Lorentz transformations corresponds to O$(3)$, and therefore the classification should be made by considering this little group with the exact polarization expressions.

\acknowledgments

Y.-H. Hyun, Y. Kim, and S. Lee are supported by Basic Science Research Program through the National Research Foundation of Korea (NRF) funded by the Ministry of Science, ICT and Future Planning (Grant No. NRF-2018R1D1A1B07049514, NRF-2016R1D1A1B03931090, and NRF-2017R1A2B4011168, respectively).\\


\end{document}